\documentclass[twocolumn,showpacs,groupedaddress,amssymb, aps, prl, nofootinbib]{revtex4}

\usepackage{graphicx}
\usepackage{bm}
\usepackage{textcomp}

\begin{document}

\title{Electromechanical effect in complex plasmas}

\author{S.~Zhdanov$^1$}

\email{zh@mpe.mpg.de}

\author{M.~Schwabe$^1$}

\author{R.~Heidemann$^1$}

\author{H.M.~Thomas$^1$}

\author{M.~Rubin-Zuzic$^1$}

\author{H.~Rothermel$^1$}

\author{T.~Hagl$^1$}

\author{A.V. Ivlev$^1$}

\author{G.E. Morfill$^1$}

\address{$^1$ Max-Planck-Institut f\"{u}r extraterrestrische Physik, 85741 Garching, Germany}

\author{V.I. Molotkov$^2$}

\author{A.M. Lipaev$^2$}

\author{O.F. Petrov$^2$ }

\author{V.E. Fortov$^2$ }

\address{$^2$ Institute for High Energy Densities, Russian Academy of Sciences, 125412 Moscow, Russia}

\author{ T. Reiter$^3$}

\address{$^3$ESA/EAC, Linder H\"ohe, 51147 K\"oln, Germany}

%
%
%
%
%


\date{Montag, Juli 28, 2008 at 20:26}

\begin{abstract}
Experimental results on an auto-oscillatory pattern observed in a
complex plasma are presented. The experiments are performed with an
argon plasma which is produced under microgravity conditions using a
capacitively-coupled rf discharge at low power and gas pressure. The
observed intense wave activity in the complex plasma cloud
correlates well with the low-frequency modulation of the discharge
voltage and current and is initiated by periodic void contractions.
Particle migrations forced by the waves are of long-range repulsive
and attractive character.
\end{abstract}


\maketitle%

In this letter we address the electromechanical effect and
associated self-sustained oscillatory and wave patterns observed in
complex plasmas with the PK-3~Plus setup on board the ISS
\cite{PK3Plus,Thomas2008}. These active experiments (including as an
item wave excitations \cite{Mierk2008}) have been performed recently
in a particularly wide range of plasma parameters \cite{Thomas2008}.
With a complex plasma in an active state an activation of the
\emph{particle cloud-plasma} feedback mechanism leading to
self-sustained oscillations is quite natural.

Complex plasmas are low pressure low temperature plasmas containing
microparticles. These particles can be visualized individually with
a laser beam, the light of which is scattered by the particles and
then recorded with a CCD camera. Under microgravity conditions in
experiments on board the ISS \cite{Thomas2008} or in parabolic
flight experiments \cite{Piel2006}, the recorded particle clouds are
essentially three-dimensional structures more or less homogeneous,
albeit inside commonly containing a void -- a 'visibly empty region'
free of particles \cite{Morfill1999}. (For the last decade "void
closure" was a long-standing challenge of complex plasma
investigations \cite{Lipaev2007}.)

The system \emph{'complex plasma with a void'} (CPV) is one of the
most intriguing phenomena detected in experiments with complex
plasmas: the CPV gives rise to self-excited macroscopic motions --
it sets the paradigm of a dissipative system capable of
auto-oscillations. (This stable auto-oscillating CPV should not be
confused with cyclic spatio-temporal microparticle generations
\cite{Cavarroc2008}, and the \emph{heart-beat instability}
\cite{Mikikian2007} studied in detail in dust forming plasmas.)
Depending on discharge conditions and plasma parameters, the CPV
could be kept stable, or excited externally into an oscillatory
state, which even in the presence of damping remains autonomically
excited.

There are a few useful analogies which help to identify the main
features of the self-excitation' mechanism.

An ability to self-sustain oscillations is typical for
\emph{auto-oscillation systems } with inertial self-excitation such
as the Helmholtz resonator well known in acoustics, and many others
\cite{Pierce1989}.

We associate low-frequency current and voltage self-pulsations, and
the accompanying particle oscillations in complex plasmas, with the
\emph{negative differential conductivity} (NDC--see
\cite{Hoffmann1975}). In this sense the CPV exibits properties
similar to some type of photoconductors \cite{Hoffmann1975},
semiconductors \cite{Esaki1970}, semi-metals \cite{Upit1966},
ferroelectric liquid-crystalline films \cite{Yablonskii2001}, carbon
nanotubes \cite{Anton2000}, nanocrystalline heterostructures
\cite{Chen2006} and other microelectromechanical systems
\cite{Sung2003,Mani2006}.

The non-linear dissipative compact formations in the patterns seen
in CPV remotely resemble \emph{oscillons}. These standing
undulations can be produced on the free surface of a liquid
(so-called Faraday's waves \cite{Faraday1831}), a granular medium
(in this case localized excitations can self-organize with possible
assembly into 'molecular' and 'crystalline' structures
\cite{Umbanhowar1996}), or nonlinear electrostatic oscillations on a
plasma boundary \cite{Stenflo1996}. Oscillons "feed" from external
shaking of the system and dissipation seems to be inherent for their
existence, likewise in our case. There remain still many open
questions in explaining the physical mechanism of oscillons.

Hydrodynamic (or hydrodynamic-like) systems provide other examples
of \emph{oscillatory patterns} fed by streaming in the system: a
flow-acoustic resonance \cite{Roley2002}, hydrothermal
\cite{Garnier2001} and plastic deformation flows \cite{Zuev2001},
mobile dunes \cite{Sauermann2001}, surface tension auto-oscillations
\cite{Kovalchuk1999}, oscillating domains in planar discharges
\cite{Struempel2000}, self-excited dust density waves
\cite{Mierk2007}, and many others.

The usual assumption is that auto-oscillations are maintained by a
sufficiently powerful instability allowing recirculation
(hysteresis) in phase space \cite{Koschmieder,Zieper}. By providing
investigations of micro-particle migrations at the atomistic level,
experiments with the CPV may help us to understand the fundamental
nature of inanimate auto-oscillations and the intrinsic dynamics of
highly dissipative non-linear structures.

Streaming ions may act as an activator of the instability in the CPV
(see, e.g. \cite{Mierk2007}). The formation of a void itself is
explained frequently by the counteraction of the confinement force
and the ion drag force \cite{Samsonov1999,Kretschmer}. Void
vibrations in an energized CPV are believed to be due to the heart
beat instability the free energy of which may arise from streaming
ions \cite{Mikikian2007}. There are a number of direct observations
of the interaction of streaming ions and dust particles
\cite{Kretschmer, Steinberg, Schweigert}.

Since straightforward measurements of flowing ions in complex
plasmas are not possible to perform without perturbing the particle
cloud structure \cite{APiel}, evidence might be extricated from the
direct observations of particle vibrations. There are two possible
options: (i) elucidate the long-range character of particle
vibrations; (ii) decode the patterns of the secondary wavefronts.
Fortunately, both options are realizable as has been proven by the
given experiments.

\begin{figure}[b]
\includegraphics[width=1\linewidth,bb = 14 14 292 161]{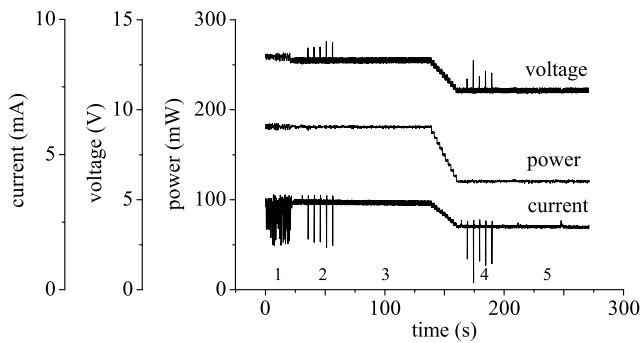}
    \caption{\label{fig:fig1} The (rms) discharge voltage, the current and the applied power
    measured during the two-stage experiment. The numbers indicate the main phases: 1 - stochastic
    stabilization,
    2 - first round of short-time pulses, 3 - stable free oscillations at a higher applied power, 4 - second round
    of short-time pulses, 5 - free oscillations at a lower applied power.
    }
\end{figure}

We performed our experiments in a PK-3~Plus chamber
\cite{PK3Plus,Thomas2008}. The parallel plate capacitively coupled
rf discharge is symmetrically driven by two electrodes, which have a
diameter of 6~cm and are separated by 3~cm (measured voltage
asymmetry does not exceed $\simeq$2\%; all measured electrotechnical
values shown below are arithmetic means). The electrodes are
surrounded by a grounded ring of 9~cm diameter and 1.5~cm width.
(More technical details of the setup can be found in
\cite{Thomas2008}). Particle vibrations were recorded at a rate 50
fps and a spatial resolution of 0.41 Mpixel at 45.05~\textmu m/pixel
(49.6~\textmu m/pixel) in the vertical (horizontal) direction. The
sample rate of the low-frequency electrotechnical measurements was
10~Hz at the stage of stable vibrations.

\begin{figure}[t]
       \includegraphics[width=0.9\linewidth,bb = 1 19 517 339]{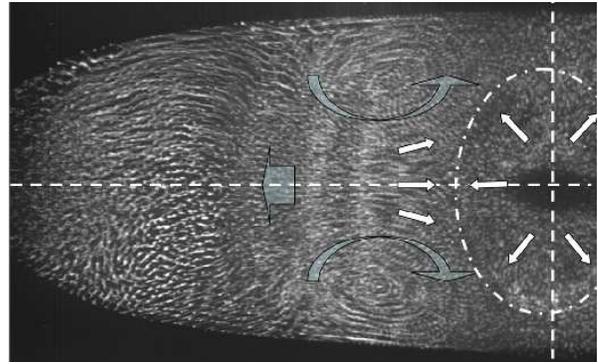}
       \caption{\label{fig:fig2} 10 superimposed  images
    shifted by one period in time are shown to reveal the global dynamical pattern of the CPV.
    The main elements of this 'dynamo-machine' are: a void (the dark elliptic-shaped
    area to the right, at this active stage open), a quasi-spherical halo (highlighted
    by the dash-dotted line) with concentric
    waves spreading around the vibrating void, two horizontal
    counter-rotating global vortices with angular velocity $\simeq$
    0.2~s$^{-1}$, and horizontal radius $\simeq$7.5~mm, and an edge 'buffer' zone
    to the left. The boundaries of the vortices form a 'waveguide'
    for oscillons in-between (oscillons are identified in the middle as a few brighter
    vertical stripes). For the given half-cycle the dominant
    particle motion is as indicated by the arrows.
    The semi-transparent arrows indicate general particle drifts. The dashed
    lines cross at the position of the void center.
    The field of view is 17.6$\times$34.7~mm$^2$. The illuminating laser sheet FWHM is about 80~$\mu$m.}
\end{figure}

\begin{figure}[t]
\includegraphics[width=0.95\linewidth,bb = 14 15 321 235]{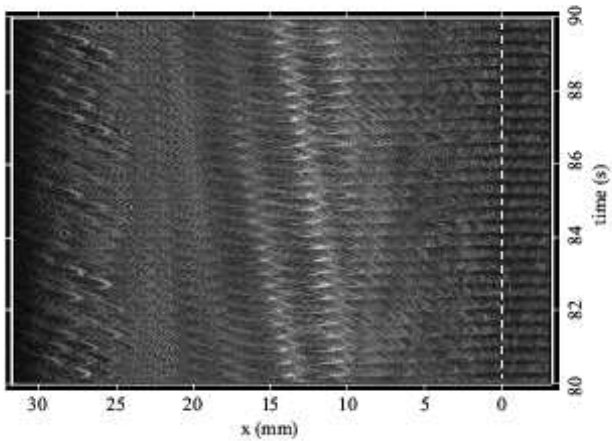}
    \caption{\label{fig:fig3} Periodgram showing horizontal oscillations of the cloud.
    The brighter spikes in
    the middle (indicating enhanced
    particle density) are oscillons (see also Fig.\ref{fig:fig2}) slowly propagating
    towards the outer edge of the cloud (to the left)
    away from the pulsating void (the horizontal
    periodic dark stripes to the right) at an approximately constant speed of 0.4~mm/s.
    The dashed line indicates the position of the void
    center. The life-time of the oscillons is $\simeq$20~s, i.e. about 200 damping times.
    Oscillons are "feed" by the CPV oscillatory energy.
    The faster edge wave-ridges (see \cite{Mierk2008}) are also clearly seen at x$>$25~mm.}
\end{figure}

The experiment, we address here, was performed in Argon at a
pressure of 9~Pa and was arranged in two stages
(Fig.\ref{fig:fig1}). In the first stage the discharge was ignited
with a peak-to-peak voltage of 37~V at an applied (rms) power of
0.181~W (the discharge power factor for the entire circuitry was
estimated as about 45-60\%). Melamine-formaldehyde particles with a
diameter of 9.2 $\pm$ 1\% \textmu m and a mass density of
1.51~g/cm$^3$ were inserted into the chamber. \footnote {This is one
of the main differenses of this experiment compared to
\cite{Mikikian2007}: the particles are injected, not grown in a
plasma, and the particle size is larger (at least one order of
magnitude).} They formed a cloud stretched horizontally (the aspect
ratio width/height $\equiv$ D/H~$\approx$~64~mm/15~mm) with a
visually pulsating elliptically-shaped void (see Fig.\ref{fig:fig2};
in the maximal "stretched" phase the void is $\sim$
7$\times$3~mm$^2$). This discharge regime allows us to observe
stable oscillations at a frequency 3-15~Hz and to perform
statistically relevant measurements during $\simeq$ 150~s. The
estimated gas damping rate, attenuating particle motion, was
10.7~s$^{-1}$. Unlike \cite{Samsonov1999,Mikikian2007,Cavarroc2008}
no contaminating (sputtering) components affected the discharge.

Next the applied power was lowered to 0.12~W, and another form of
the heart-beat instability with its almost irregular large-amplitude
void constrictions started. The stable oscillation phase we address
here would seem to be a completely different phenomenon. Details on
the unstable (heart beat) phase will be published elsewhere.

The experiment started out from the stochastically stabilized
complex plasma (t=0 in Fig.\ref{fig:fig1}; for details of stochastic
stabilization see \cite{Thomas2008}). At 20~s the external
stabilization was turned off and after $\Delta t \simeq$1~s delay
the auto-oscillations appeared. Next at 31~s the plasma was
stimulated by a series of six short-time voltage pulses produced by
the function generator. The pulses, with a negative amplitude of
-50~V, were applied to the bottom electrode through a 20kOhm feed
resistor. This produces short-time DC voltage shifts of a few volts
which leads to shock compression of the particle cloud. This
technique can also be used as a stability test: The CPV reverted
rapidly to its former stable condition after each applied pulse.
After this the CPV was observed to freely oscillate for $\simeq$83~s
without external forcing.

Fig.\ref{fig:fig2} visualizes the dynamical activity. Surprisingly
the "shaking" divides the particle cloud into two counter-moving
parts which forms a stagnation zone with nearly zero particle
velocity at the interface. Outside the stagnation zone the particles
are seen to move along the axis at nearly constant velocity (in this
particular half cycle) as if they were \emph{attracted} by the void.
Apparently, in this phase the void behaves like a negatively charged
probe, and long range attraction is due to the ion collection effect
\cite{Samsonov2001}. (In the next half a cycle, as the void tends to
close, the particles are \emph{repelled} from the center). Simple
estimates based on particle behavior analysis show a rather weak
plasma charge oscillation with maximal decompensation of the order
of $\delta n/n=5\cdot10^{-3}$.

The periodic auto-oscillation pattern formed by particle horizontal
vibrations is shown in Fig.\ref{fig:fig3}. For depicting the
oscillation pattern, we follow a simple procedure proposed in
\cite{Mierk2007}. From each image of the recorded sequence a narrow
slab of size 35x5~mm$^2$ centered over the void vertical position is
extracted. Then the slab is 'compressed' into a line by adding up
the pixel intensities perpendicular to its longer side. The result
is plotted (as a periodgram) for every frame as shown in
Fig.\ref{fig:fig3}, forming a (x, t) map. The brighter regions of
this map correspond to higher particle densities. The darkest
regions are particle-free, that means that the void is maximally
open. The fundamental oscillation frequency shown by the cloud is
\begin{equation}\label{eq:frequency}
f_{osc}=2.81\pm 0.03 ~Hz.
\end{equation}

\begin{figure}[b]
   \includegraphics[width=1.0\linewidth, bb = 15 18 185 227]{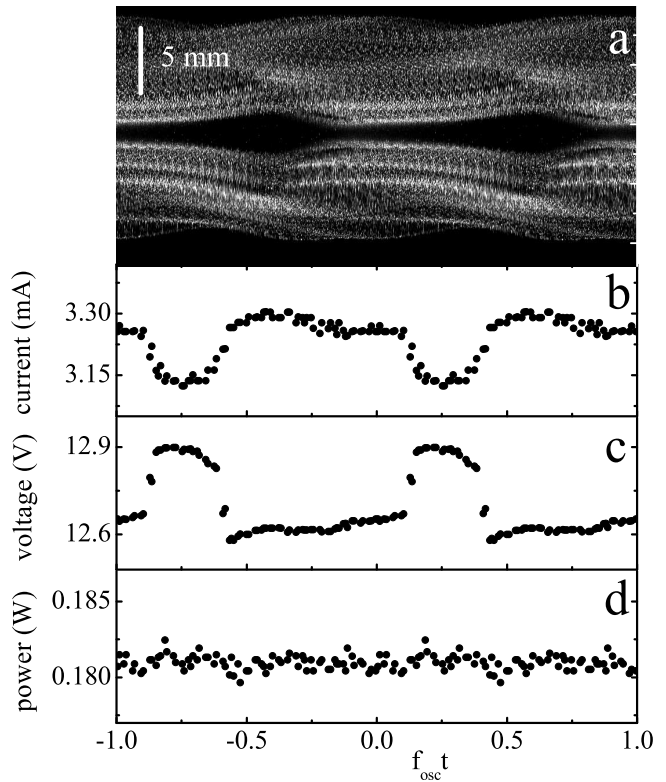}
    \caption{\label{fig:fig4}(a) Vertical oscillations of the
    cloud, (b) the (rms) rf current, (c) rf voltage and (d) applied (forward) rf power
    vs. reduced time ($\equiv f_{osc}t$). Two 'reconstructed' oscillation periods are shown.
    Reconstruction, based on superposition of the data points taken over 20
    consecutive periods (phase 3 in Fig.\ref{fig:fig1}), demonstrates clearly local quasi-periodicity of the
    discharge electrical signals (periodicity time is limited by
    slowly drifting oscillons). Lack of any periodicity in the applied power signal is also evident.
    (The forward power is held constant by a servo control loop inside the rf generator.) }
\end{figure}

Fig.\ref{fig:fig4} shows a correlation between vertical cloud
oscillations and all electrical signals. There is no correlation
with the applied power. To depict the pattern of vertical
oscillations the aforementioned procedure is carried out over a
narrow vertical slab of size 2x20~mm$^2$. The result is shown in
Fig.\ref{fig:fig4}a as a superposition of consecutive periods.

Comparison with the oscillatory components of the effective voltage
and current (Fig.\ref{fig:fig4}b,c) show an almost \emph{anti-phase}
behaviour that is typical to pattern-creating gas discharges (see,
e.g. \cite{Danijela2004}). Mean components of these signals
(corrected to measurement offset errors) are
\begin{equation}\label{eq:current imbalance}
              \nonumber
                 <I>=3.24\pm 0.03 ~mA, ~<U>=12.70\pm 0.02 ~V.
                 \end{equation}
From this we obtain an average ohmic discharge resistance of
$\simeq$4~kOhm. (Without particles the plasma resistance was
$\simeq$3.4~kOhm. This value agrees well with that estimated from
\cite{Godyak1991}.)

Note that sharp current minima (voltage maxima) are apparently
linked to the vertical cloud compression, whereas the void opening
relates to a weak rise in the current. All this suggests an
association with the variation of the discharge capacitance -- an
"electromechanical effect". In this sense the CPV might resemble a
\emph{varicap}, a device whose capacitance varies under the applied
voltage.

Based on this analogy and using an equivalent circuit model
\cite{Godyak1991} it is easy to show that the expected variation of
the rms current is of the order of the
\begin{equation}
\frac{\delta I}{<I>}\simeq-\eta\frac{C_{cloud}}{C},
~\eta=\frac{\omega^2\tau^2}{1+\omega^2\tau^2}, ~\tau=RC,
\end{equation}
where R is the discharge resistance , C the discharge capacitance,
C$_{cloud}$ the CPV capacitance. Assuming $\tau\simeq$1 we estimate
C$\simeq$3~pF. Hence, the observed oscillation amplitude would be
explained if C$_{cloud}\simeq$0.4~pF, which is quite reasonable.

Note also that the active oscillatory state of the CPV is
accompanied by a periodic pulsation in the discharge glow as well.

Extrapolating probe measurements \cite{Klindworth2006} we estimate
the plasma parameters as $n_e\sim10^8$~cm$^{-3}$, $T_e\sim$2 --
3~eV. Interparticle separation averaged over the entire cloud area
(Fig.\ref{fig:fig2}) is $<\Delta>\simeq$230~$\mu$m. The highest
compression occurs at the spikes (Fig.\ref{fig:fig3}),
$\Delta_{min}=$173$\pm$16~$\mu$m. Outside the spikes compression is
less, $<\Delta>$=300-350~$\mu$m. At the kinetic level we see that
particles are first accelerated to high speeds
($v_{max}$=18.9$\pm$0.4~mm/s), then they are decelerated forming
spikes -- oscillons which are clearly seen in Fig.\ref{fig:fig2} as
vertically elongated constrictions. Following \cite{Mierk2007}, we
estimate the particle charge $Z_d$$\sim$9000e, the dust sound speed
C$_{DAW}$$\sim$6$-$7~mm/s, and the dust plasma frequency
$f_{dust}=\omega _{dust}/2\pi \approx$20~Hz.

The origin of the proposed electromechanical effect could be due to
the cloud stretching, multiplication or selective harmonic
amplification of coupled oscillations of the particles and the
electrical circuitry feeding the discharge. In the case studied here
it is a self-sustained low frequency resonant oscillator. We
conclude that the self-excitation leading to the regular repeatable
constrictions of the void in the particle cloud is due to the free
energy in plasma ions drifting relative to the microparticles.

\begin{acknowledgments}
    The authors would like to acknowledge valuable discussions with
    R.~S$\ddot{u}$tterlin, P.~Huber, U.~Konopka, and all members of the PK-3~Plus team.
This work was supported by DLR/BMWi Grant No. 50WP0203, and by RFBR
Grant No. 06-02-08100.
\end{acknowledgments}

\end{document}